\documentclass[preprint, prb]{revtex4}

\usepackage{amsmath}

\newcommand{{\bff}}{\boldsymbol{f}}

\newcommand{\gradv}{\boldsymbol{\nabla}}

\def\v#1{{\bf#1}}

\begin{document}

\title{Reply to ``Comment on `Can Maxwell's equations be obtained from the continuity equation?",' by E.\ Kapu\`scik [Am. J. Phys. {\bf 77}, 754 (2009)]}

\author{Jos\'e A. Heras}
\affiliation{Departamento de Ciencias B\'asicas, Universidad Aut\'onoma Metropolitana, Unidad Azcapotzalco, Av. San Pablo No. 180, Col. Reynosa, 02200, M\'exico D. F. M\'exico and Departamento de F\'isica y Matem\'aticas, Universidad Iberoamericana, Prolongaci\'on Paseo de la Reforma 880, M\'exico D. F. 01210, M\'exico}

\maketitle

\noindent

Kapu\`scik's comments\cite{1} on my derivation of Maxwell's equations from the continuity equation\cite{2} are based on the idea that I neither used relativistic notation nor considered material media. My derivation has also been commented on by Jefimenko.\cite{3,4} Kapu\`scik's objections provide the opportunity to clarify certain important points not emphasized in my original paper.\cite{2} 

Kapu\`scik invokes the well-known result that given the continuity equation in spacetime
\begin{equation}
\label{1}
\partial_\nu J^\nu=0,
\end{equation}
we can write the solution of this equation as
\begin{equation}
J^{\nu}=\partial_\mu H^{\mu\nu},
\end{equation}
where $H^{\mu\nu}$ is an antisymmetric tensor.\cite{5} Kapu\`scik points out that for fixed $J^\nu=[c\rho, (\v J)^j]$, Eq.~(2) becomes a differential equation for $H^{\mu\nu}$ that is equivalent to the inhomogeneous Maxwell equations. He concludes that the correct statement should be that the inhomogeneous Maxwell equations follow from continuity equation. However, on the basis only of Eq.~(2) with $J^\nu=[c\rho, (\v J)^j]$, the tensor $H^{\mu\nu}$ is not necessarily the electromagnetic tensor field, that is, Eq.~(2) does not necessarily represent the inhomogeneous Maxwell equations. In (1+3) notation, Eq.~(2) implies that
\begin{equation}
[c\rho, (\v J)^j]=(\partial_{i}H^{i0}, \; \partial_{0}H^{0 j}+\partial_{i}H^{ij}).
\end{equation}
But we are free to specify the components $H^{i0}$ and $H^{ij}$. For sources in vacuum, only the specific choice (Gaussian units):
\begin{equation}
H^{i0}=\frac{c}{4\pi}(\v E)^{i}\quad {\rm and}\quad H^{ij}=-\frac{c}{4\pi}\varepsilon^{ijk}(\v B)_k,
\end{equation}
allows us to identify Eq.~(3) with the equations:
\begin{equation}
\gradv\cdot\v E=4\pi\rho\quad{\rm and}\quad \gradv\times \v B-\frac{1}{c}\frac{\partial \v E}{\partial t}=\frac{4\pi}{c}\v J.
\end{equation}
For different choices of the components $H^{i0}$ and $H^{ij}$, Eq.~(3) cannot be identified with Eq.~(5).\cite{6} Furthermore, we cannot yet claim that Eq.~(5) represents the inhomogeneous Maxwell's equations because $\v E$ and $\v B$ in Eq.~(5) are not necessarily electric and magnetic fields.\cite{7} To be sure that $\v E$ and $\v B$ in Eq.~(5) are electric and magnetic fields the quantities $\gradv\cdot\v B$ and $\gradv\times \v E+(1/c)\partial \v B/\partial t$ need to be specified as
\begin{equation}
\gradv\cdot\v B=0\quad{\rm and}\quad \gradv\times \v E+\frac{1}{c}\frac{\partial \v B}{\partial t}=0,
\end{equation}
or equivalently as $\partial_\mu{\!^*}\!H^{\mu\nu}=0$, but this last relation is not mentioned in Ref.~\onlinecite{1} (the dual of $H^{\mu\nu}$ is defined as $^*\!H^{\mu\nu}=(1/2)\varepsilon^{\mu\nu\kappa\sigma}H_{\kappa\sigma}$). In other words Eq.~(2) cannot yet be identified with the inhomogeneous Maxwell equations because $H^{\mu\nu}$ is not completely determined by Eq.~(2).

From the Helmholtz theorem for antisymmetric tensors\cite {8} a tensor $H^{\mu\nu}$ that vanishes at infinity is completely determined by specifying its divergence $\partial_\mu H^{\mu\nu}$ and the divergence of its dual $\partial_\mu{\!^*}\!H^{\mu\nu}$. Equation~(2) of Ref.~\onlinecite{1} specifies $\partial_\mu H^{\mu\nu}$, but $\partial_\mu{\!^*}\!H^{\mu\nu}$ has not been specified. If we maliciously specify $\partial_\mu{\!^*}\!H^{\mu\nu}$ by the relation $\partial_\mu{\!^*} H^{\mu\nu}= S^{\nu}$, where $S^{\nu}=(\gradv\cdot\v s, \gradv \times\v s-(1/c)\partial \v s/\partial t)$ with $\v s$ being an arbitrary vector (note that $\partial_\nu S^{\nu}=0$), then the fields $\v E$ and $\v B$ in Eq.~(5) are not Maxwell's fields and therefore Eq.~(2) cannot be identified with the inhomogeneous Maxwell equations.

Kapu\`scik's argument has been explored on the basis of the de Rham theorem in differential forms,\cite{9} which implies that given $J^\nu$ satisfying $\partial_\nu J^\nu =0$, there exists an antisymmetric tensor $F^{\mu\nu}$ such that $J^\nu=\partial_\mu F^{\mu\nu}$. This $F^{\mu\nu}$ is not uniquely determined. We can add to it the antisymmetric tensor $\varepsilon^{\mu\nu\alpha\beta}\partial_{\alpha}\chi_\beta$, where $\chi_\beta$ is an arbitrary four-vector, and the relation $J^\nu=\partial_\mu F^{\mu\nu}$ does not change. This ambiguity in the definition of $F^{\mu\nu}$ prevents us from giving a physical meaning to this field. From the invariance of $\partial_\mu F^{\mu\nu}=J^\nu$ under the field transformation $F^{\mu\nu}\to \widetilde{F}^{\mu\nu}= F^{\mu\nu}+ \varepsilon^{\mu\nu\alpha\beta}\partial_{\alpha}\chi_\beta$, that is, $\partial_\mu \widetilde{F}^{\mu\nu}=J^\nu$, we may explore the idea of fixing the function $\chi_\beta$ so that $\partial_\mu{\!^*}\!\widetilde{F}^{\mu\nu}=0$. Suppose that $F^{\mu\nu}$ satisfies $\partial_\mu F^{\mu\nu} = J^\nu$, but not $\partial_\mu{\!^*}\!F^{\mu\nu}=0$. This means that $\partial_\mu{\!^*}\!F^{\mu\nu}=g^\nu \neq 0$. Then we make a field transformation to $F^{\mu\nu}$ and require that $\widetilde{F}^{\mu\nu}$ satisfies $\partial_\mu{\!^*}\widetilde{F}^{\mu\nu}=0$, that is, $\partial_\mu{\!^*}\!\widetilde{F}^{\mu\nu}=0= g^\nu+ \partial_\mu\partial^\mu\chi^\nu -\partial^\nu\partial_\mu\chi^\mu$. The result $\partial_\mu{\!^*}\!\widetilde{F}^{\mu\nu}=0$ would be possible only if we could find the solution of $\partial_\mu\partial^\mu\chi^\nu - \partial^\nu\partial_\mu\chi^\mu=-g^\nu$. Such a solution is not known and so we cannot guarantee the existence of the function $\chi^\mu$, and thus the relation $\partial_\mu{\!^*}\!\widetilde{F}^{\mu\nu}=0$ cannot generally be established. The derivation of Maxwell's equations based only on the de Rahm theorem turns out to be unsatisfactory and further additional assumptions are required.

Kapu\`scik also claims that my derivation that the homogeneous Maxwell equations follow from the continuity equation must be considered to be incorrect because the homogeneous Maxwell equations are based on a tensor field $F^{\mu\nu}$ that is conceptually independent of the sources of the electromagnetic field present in the inhomogeneous Maxwell equations. In my derivation I considered only sources in vacuum satisfying the continuity equation and therefore $\v E$ and $\v B$ obtained in this way (or equivalently $F^{\mu\nu}$) are causally generated by the sources $\rho$ and $\v J$ (or equivalently $J^{\nu}$) as is seen in Eq.~(28) of Ref.~\onlinecite{2}. That is, in the context of my derivation the fields $\v E$ and $\v B$ (or equivalently $F^{\mu\nu}$) in the homogeneous Maxwell equations depend on the sources present in the inhomogeneous Maxwell equations.

Kapu\`scik claims that my derivation of Maxwell's equations applies only in vacuum and that Maxwell equations apply to arbitrary media. Equation~(6) holds in material media, but Eq.~(5) must be replaced by
\begin{equation}
\gradv\cdot\v D=4\pi\rho\quad{\rm and}\quad \gradv\times \v H-\frac{1}{c}\frac{\partial \v D}{\partial t}=\frac{4\pi}{c}\v J,
\end{equation}
where $\v D$ and $\v H$ are related to $\v E$ and $\v B$ through the polarization $\v P$ and the magnetization $\v M$ of the material media by
\begin{equation}
\v D=\v E+4\pi\v P \quad{\rm and}\quad \v H=\v B-4\pi\v M.
\end{equation}
I agree with Griffiths who has written:\cite{10} ``Some people regard \ldots [Eqs.~(6) and (7)] as the ``true" Maxwell's equations, but please understand that they are in no way more ``general" than \ldots [Eqs.~(5) and (6)]; they simply reflect a convenient division of charge and current into free and nonfree parts." From Eqs.~(7) and (8) we obtain
\begin{align}
\gradv\cdot\v E&=4\pi(\rho-\gradv\cdot\v P), \label{9}\\
\gradv\times \v B-\frac{1}{c}\frac{\partial \v E}{\partial t} &=\frac{4\pi}{c}\bigg(\v J +c\gradv\times \v M+\frac{\partial \v P}{\partial t}\bigg). \label{10}
\end{align}
The right-hand sides of Eqs.~\eqref{9} and \eqref{10} display the division of charge and current into free and nonfree parts. The derivation of Maxwell's equation given for free sources in Ref.~\onlinecite{2} can be generalized to free and nonfree sources\cite{10} by making the substitutions $\rho\to\rho_{\rm T}$ and $\v J\to\v J_{\rm T}$, where $\rho_{\rm T}$ and $\v J_{\rm T}$ are the total sources defined by
\begin{equation}
\rho_{\rm T} =\rho_f -\gradv\cdot\v P \quad{\rm and}\quad \v J_{\rm T}=\v J_f+ c\gradv\times \v M+\frac{\partial\v P}{\partial t},
\end{equation}
where $\rho_f$ and $\v J_f$ are free parts and $-\gradv\cdot\v P$ and $c\gradv\times \v M+\partial\v P/\partial t$ are nonfree parts. Therefore the derivation of Maxwell's equations from the continuity equation applies not only to the vacuum but also to material media.

Kapu\`scik claims that the homogeneous Maxwell's equations
\begin{equation}
\partial_\mu F_{\nu\lambda}+\partial_\lambda F_{\mu\nu}+\partial_\nu F_{\lambda\mu}=0,
\end{equation}
can be obtained from the equation $\partial_\mu F_{\nu\lambda}+\partial_\lambda F_{\mu\nu}+\partial_\nu F_{\lambda\mu}=J_{\mu \nu\lambda}$, because the Faraday induction law leads to $J_{\mu \nu \lambda }( x) =0$. Unfortunately, his argument is circular. It considers the Faraday induction law, $\gradv\times \v E+ (1/c)\partial \v B/\partial t=0$, as one of Maxwell's equations to obtain Eq.~(12). But it does not make sense to assume one of Maxwell's equations to derive Maxwell's equations themselves.

Kapu\`scik claims that my result should be treated as finding a particular solution of the vacuum Maxwell equations rather than proving that Maxwell equations for arbitrary media follow from the continuity equation. As noted, Maxwell's equations in matter can also be obtained from the continuity equation. I agree with Kapu\`scik's comment that my derivation involves only retarded fields and that Maxwell's equations admit also advanced fields. However, we can derive Maxwell's equations from the continuity equation using the free-space Green function of the wave equation, which admits retarded and advanced forms. The fields $\v E$ and $\v B$ obtained in such a derivation would be defined in terms of this Green function and would represent either retarded or advanced fields when the retarded or advanced form of the Green function is made explicit.

I emphasize that the derivation of Maxwell's equations from the continuity equation formally arises from applying an existence theorem. If the proof of the theorem is incorrect or the mentioned application is inconsistent, then the derivation could be questioned. Kapu\`scik\cite{1} has not questioned the validity of the theorem.

\begin{acknowledgments}
The author thanks Edward Kapu\`scik for useful comments. The support of the Fondo UIA-FICSAC is gratefully acknowledged.
\end{acknowledgments}


\begin{thebibliography}{}
\bibitem{5}Greek indices $\mu, \nu, \kappa \ldots$ run from 0 to 3; Latin indices $i,j,k,\ldots$ run from 1 to 3; $x=x^{\mu}=(ct,\v x)$ is the field point and $x'=x'^{\mu}=(ct',\v x')$ the source point; the signature of the metric of spacetime is $(+,-,-,-);$ $\varepsilon^{\mu\nu\alpha\beta}$ is the totally antisymmetric four-dimensional tensor with $\varepsilon^{0123}=1$ and $\varepsilon^{ijk}$ is the totally antisymmetric three-dimensional tensor with $\varepsilon^{123} = 1$. Summation convention on repeated indices is adopted. A four-vector is represented in the (1+3) notation as $F_\nu=(f_0,\v F)$.
\bibitem{6}We could make the choice $H^{i0}=-[c/(4\pi)](\v E)^{i}$ and $H^{ij}=-[c/(4\pi)]\varepsilon^{ijk}(\v B)_k, $ for which Eq.~(3) does not yield Eq.~(5).
\end{thebibliography}
\end{document}